\documentclass[prl,aps,amsmath,twocolumn,amsfonts,showpacs,superscriptaddress,floatfix]{revtex4}

\usepackage{bm}
\usepackage{graphicx}
\usepackage{color}

\def\be{\begin{equation}}
\def\ee{\end{equation}}

\def\Tr{\mathop{\rm Tr}}

\def\eps{\varepsilon}

\begin{document}

\title{
Ferromagnetic Josephson junction with precessing magnetization
}

\author{Manuel Houzet}
\affiliation{INAC/SPSMS, CEA Grenoble, 17 rue des Martyrs, 38054 Grenoble Cedex, France}

\date{\today}

\pacs{74.45.+c, 76.50.+g.}

\begin{abstract}
The Josephson current in a diffusive superconductor/ferromagnet/superconductor junction with precessing magnetization is calculated within the quasiclassical theory of superconductivity. When the junction is phase-biased, a stationary current (without a.c. component) can flow through it despite the non-equilibrium condition. A large critical current is predicted due to a dynamically induced long range triplet proximity effect. Such effect could be observed in a conventional hybrid device close to the ferromagnetic resonance.
\end{abstract}

\maketitle

The proximity effect in a ferromagnetic (F) metal in contact with a conventional superconductor (S) is usually short ranged \cite{buzdin}. Indeed, the conversion from Cooper pairs in the superconductor to Andreev pairs in the ferromagnet involves two electrons with opposite spins. They get quickly dephased due the large value of the exchange field acting on their spins. Yet, it was predicted that a long range proximity effect can take place when the magnetization rotates spatially \cite{bergeret-rmp} or when spin-flip processes take place \cite{eschrig} in vicinity of the F/S interface. Indeed, Andreev pairs involving electrons with parallel spins are then created and can propagate on a long distance. The long range proximity effect was invoked to explain the large Josephson current measured through half-metallic (HM) chromium oxide \cite{keizer} as well as superconducting phase-periodic conductance oscillations in an Andreev interferometer made of a long metallic wire with helimagnetic order \cite{sosnin}. However, the crossover from short range to long range proximity effect in the same device remains to be observed. Recently, it was suggested to measure the strong variations of the Josephson current through a trilayer ferromagnetic junction by tuning the relative orientations of the magnetizations \cite{houzet-buzdin}.

\begin{figure}
\begin{center}
\includegraphics*[width=0.7 \columnwidth]{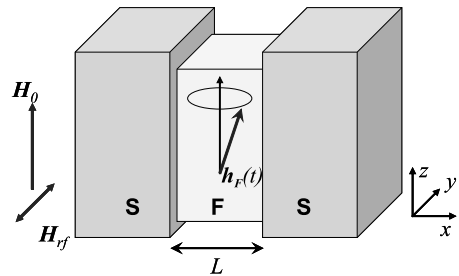}
\caption{
Geometry of the ferromagnetic Josephson junction with precessing magnetization and configuration of d.c. and r.f. magnetic fields inducing the ferromagnetic resonance.
}
\label{fig1}
\end{center}
\end{figure}

In this article, we propose that the long range triplet proximity effect can be stimulated by varying {\it in time} (rather than {\it in space}) the orientation of the magnetization in the ferromagnet. Specifically, we consider a diffusive metallic S/F/S junction with dynamical precession of the magnetization, as shown in fig.~\ref{fig1}. In the ferromagnet, the conduction electrons feel the time-varying exchange field:
\be
\label{eq:prec}
\bm{h}_F(t)=h_F(\sin\theta\cos\Omega t,\sin\theta\sin\Omega t,\cos\theta)
\ee 
which is proportional to the magnetization and assumed to be spatially uniform. Here, $\Omega/2\pi$ is the precession frequency around $\hat z$-axis, $\theta$ is a tilt angle, and the amplitude of exchange field, $h_F$, is constant at temperatures well below the Curie temperature. In the rotating frame, such precession can be viewed as a difference between spin-resolved chemical potentials in the leads, $\bm{\mu}_s=\hbar\Omega\hat{z}$. At finite tilt angle, $\bm{h}_F$ has a component transverse to $\bm{\mu}_s$. Thus, it creates a non-equilibrium situation for the conduction electrons. In particular, the device could act as a spin pump when the leads are in the normal state \cite{brataas}. Moreover, in the superconducting state, the non colinear orientation of $\bm{h}_F$ and $\bm{\mu}_s$ generates a long range proximity effect similar to the case studied in Ref.~\cite{houzet-buzdin}. Despite the non-equilibrium condition, we find that a stationary (d.c. only) Josephson current flows through the junction when a phase difference is applied to the leads. In a long junction, its amplitude is comparable to that of a normal (N) metallic Josephson junction with the same length, provided that $\hbar\Omega$ is comparable with $k_BT_c$ ($T_c$ is the superconducting critical temperature of the leads) and $\theta$ is large enough, while it would be exponentially suppressed in the absence of precession.

Recently, a Josephson effect was also predicted in S/HM/S junctions when magnetic impurities located at the interfaces dynamically precess and mediate spin-flip tunneling processes for the conduction electrons \cite{maekawa}. It was also suggested that this effect could be observed at the ferromagnetic resonance (FMR). In the present work, we address the opposite case when the Fermi energy much exceeds the exchange field and the spin polarization in the ferromagnet is small. When the junction's length, $L$, much exceeds the ferromagnetic coherence length, $\xi_F=\sqrt{\hbar D/h_F}$, where $D$ is the diffusion constant, we will show that the Josephson current has two dominant contributions. The first one comes from long range triplet proximity effect. The second one arises from the interference between short range and long range proximity effects out of equilibrium and it displays a non-analytical temperature dependence close to $T_c$. Both have an oscillatory behaviour, depending on the ratios between $\hbar\Omega$, $k_BT_c$, and the Thouless energy $E_T=\hbar D/L^2$. This provides a mechanism for $\pi$-coupling \cite{buzdin} in long ferromagnetic Josephson junctions. In the following, we derive both components, discuss their properties and the conditions to observe them in conventional F/S structures.

Within the quasiclassical theory of superconductivity, the current flowing through the junction is \cite{usadel,larkin}:
\be
I(t)=-\frac{\pi G L}{8e}\Tr [ \tau_z (\check{g} \circ \nabla \check{g})^K(t,t)],
\label{eq:courant1}
\ee
where the quasiclassical Green's function $\check{g}(x,t,t')$ is a matrix in spin, Nambu, and Keldysh spaces and it obeys the Usadel equation along the ferromagnetic layer:
\be
\label{eq:usadel}
-i D \nabla( \check{g}\circ \nabla \check{g})+ [(i \partial_t+\bm{h}_F(t).\bm{\sigma})\tau_z \delta(t-t_1)\stackrel{\circ}{,} \check{g}]=0.
\ee
Here, $G$ is the conductance of the ferromagnetic layer, the spatial derivative is taken along $\hat{x}$-axis, $e$ is the elementary charge, $\sigma_i$ and $\tau_j$ ($i,j=x,y,z$) are the Pauli matrices in spin and Nambu spaces, respectively, and $\circ$ denotes the time convolution. (Units with $\hbar=k_B=1$ are adopted from now.) Note also that the orbital effect generated by the magnetization has been neglected in Eq.~(\ref{eq:usadel}) as it is usually done \cite{buzdin}. Moreover, $\check{g}$ obeys the properties: $\Tr \check{g}=0$ and $\check{g}\circ \check{g}=1$, and it has a triangular structure in Keldysh space with the retarded, advanced and Keldysh components: $\hat{g}^R$, $\hat{g}^A$, and $\hat{g}^K$, respectively. The normalisation condition is fulfilled provided
$\hat{g}^{R/A}\circ\hat{g}^{R/A}=1$ and
$\hat{g}^K=\hat{g}^R\circ \hat{f}-\hat{f} \circ \hat{g}^A,$
where $\hat{f}$ is a distribution function matrix which obeys $[\hat{f},\tau_z]=0$ \cite{larkin}.

In the present study, we assume that there is a good electric contact at the F/S interfaces and we neglect the inverse proximity effect in the leads. This yields the boundary conditions:
\be
\label{eq:boundary}
\check{g}(x=\pm\frac{L}{2},t,t')=\check{g}_{s,\mp\chi/2}(t-t').
\ee
Here, $\chi$ is the phase difference between the leads. The quasiclassical Green's function in a superconductor with phase $\phi$ is defined in energy space by
$\hat{g}^{R/A}_{s,\phi}(\eps)=
[-i\eps\tau_z+\Delta\tau_\phi/\sqrt{\Delta^2-(\eps \pm i \Gamma)^2}]$ 
and $\hat{g}^{K}_{s,\phi}=(\check{g}^R_{s,\phi}-\check{g}^A_{s,\phi})f_T$, where $\tau_\phi=\cos\phi\tau_x-\sin\phi\tau_y$, $f_T(\eps)=\tanh(\eps/2T)$, and $\Delta$ is the conventional superconducting gap at temperature $T$. We also introduced a small, phenomenological depairing parameter $\Gamma$ which may account for the current-induced depairing, as well as scattering on magnetic impurities in the leads.

In order to determine the current through the junction, one has to solve Eqs.~(\ref{eq:usadel})-(\ref{eq:boundary}). Despite a time-varying exchange field in Usadel equation~(\ref{eq:usadel}), the problem is equivalent to a stationary (though non-equilibrium) one. Indeed, one may perform the unitary transformation: 
\be
\check{g}(t,t')\rightarrow V U(t) \check{g}(t,t') U(t')^\dagger  V^\dagger,
\ee 
where
$U(t)=\exp (i\Omega t \sigma_z /2)$
transforms from the laboratory frame into a rotating frame and absorbs the time-dependent terms in Usadel equation (\ref{eq:usadel}), while $V=\exp(i \alpha \sigma_y/2)$, with $\tan\alpha=h_F \sin\theta/(h_F\cos\theta+\Omega/2)$, rotates the spin quantization axis.  Specifically, $V$ aligns the effective exchange field with amplitude $J=[h_F^2\sin^2\theta+(h_F\cos\theta+\Omega/2]^{1/2}$ along $\hat{z}$-axis and rotates the precession axis away from it. Eq.~(\ref{eq:usadel}) now takes a simple form in energy space:
\be
\label{eq:usadel2}
-i D \nabla( \check{g}(x,\eps) \nabla \check{g}(x,\eps))+ [(\eps+J \sigma_z)\tau_z,\check{g}(x,\eps)]=0.
\ee
The boundary conditions (\ref{eq:boundary}) yield:
$\check{g}(x=\pm\frac{L}{2},\eps)= V \check{g}_{s,\mp\chi/2}(\eps+\Omega\sigma_z/2) V^\dagger$
and the current (\ref{eq:courant1}) takes the form
\be
\label{eq:current}
I=-\frac{ G L}{16e}\int d\eps \Tr [ \tau_z (\check{g} \nabla \check{g})^K],
\ee
Let us emphasize that the special time-dependence in Eq.~(\ref{eq:prec}) is crucial for the non-equilibrium problem (\ref{eq:usadel}) to be formulated as a stationary one, see Eq.~(\ref{eq:usadel2}). For a different time-dependence and/or for applied bias voltage, one could use the procedure formulated in Ref. \cite{cuevas} to study numerically the conductance in S/N/S junctions. 

In spite of the above simplification, we are still dealing with a complicated non-linear equation (\ref{eq:usadel2}). Now, we assume that the temperature is close to $T_c$, so that $\Delta \propto[T_c(T_c-T)]^{1/2}$ is vanishingly small. Thus, we can look for a solution 
$\check{g}=\check{g}_0+\check{g}_1+\check{g}_2+\dots$ in the series expansion around the normal state solution, when $\Delta/T_c\ll 1$. In zeroth order, one finds $\hat{g}_0^{R/A}=\pm\tau_z$ and
\be
\hat{f}_0=f_+-f_-\left[
\sin \alpha
\left(
\sigma_x\Re\frac{\mathrm{ch} q x}{\mathrm{ch}  \frac{qL}{2}}+
\sigma_y\Im\frac{\mathrm{ch}  q x}{\mathrm{ch}   \frac{qL}{2}}\right)
-\cos\alpha\sigma_z
\right],
\ee
where $f_\pm(\eps)=[f_T(\eps+\Omega/2)\pm f_T(\eps-\Omega/2)]/2$ and $q=\sqrt{2iJ/D}$.
We note that $\hat{f}_0$ is diagonal in Nambu space and the non diagonal components in spin space only appear at finite $\Omega$. In the first order in $\Delta$, Eq.~(\ref{eq:usadel2}) yields:
\be
-iD\nabla^2\hat{g}^{R/A}_1
\pm\{\eps+J\sigma_z,\hat{g}^{R/A}_1\}=0
\ee
which is solved with $\hat{g}^{R}_1=\bar{g}^{R}_{10}+\bm{\bar{g}}^{R}_1.\bm{\sigma}$, where:
\begin{subequations}
\begin{eqnarray}
&&\bar{g}^{R}_{1x}=-F^R_- \sin\alpha
\left[
\frac{\sinh k(L/2-x)}{\sinh k L} \tau_{\chi/2}
\right.
\nonumber
\\
&&
\qquad\qquad\left.
+
\frac{\sinh k(L/2+x)}{\sinh k L} \tau_{-\chi/2}
\right],
\\
&&\bar{g}^{R}_{10}\pm\bar{g}^{R}_{1z}=
(F^R_+ \pm F^R_- \cos\alpha)
\left[
\frac{\sinh p_\pm(L/2-x)}{\sinh p_\pm L} \tau_{\chi/2}
\right.
\nonumber
\\
&&
\qquad\qquad\left.
+
\frac{\sinh p_\pm(L/2+x)}{\sinh p_\pm L} \tau_{-\chi/2}
\right],
\end{eqnarray}
\end{subequations}
and $\bar{g}^R_{1y}=0$ while $\hat{g}^{A}_1=-\tau_z(\hat{g}^{R}_1)^\dagger\tau_z$. Here, $k=\sqrt{- 2 i \eps/D}$, $p_\pm=\sqrt{-2 i (\eps \pm J)/D}$, and $F^R_\pm(\eps)=[F^R(\eps+\Omega/2)\pm F^R(\eps-\Omega/2)]/2$, where $F^R(\eps)=i\Delta/\sqrt{(\eps+i\Gamma^2)-\Delta^2}$ \cite{note}. We note that the component $\bar{g}_{1x}$ is long ranged, while $\bar{g}_{10}$ and $\bar{g}_{1z}$ are short ranged. It is straightforward to check that the Keldysh component is solved with $\hat{f_1}=0$ and that the current (\ref{eq:current}) vanishes up to the first order in $\Delta$. 

In the second order in $\Delta$, the current is:
\begin{eqnarray}
\label{eq:current2}
I&=&-\frac{ G L}{16e}\int d\eps \Tr 
\left[ 
\tau_z 
\left(
2\nabla\hat{f_2}
-\hat{g}^R_1\nabla \hat{f}_0\hat{g}^A_1
\right.
\right.
\nonumber
\\
&&\left.
\left.
+\Re\left\{
[\hat{g}^R_1\nabla \hat{g}^R_1-\nabla\hat{g}^R_1\hat{g}^R_1]\hat{f}_0
-[\hat{g}^R_1]^2\nabla \hat{f}_0
\right\}
\right)
\right]
.
\quad
\end{eqnarray}
(The terms with $\hat{g}^{R/A}_2$ have been eliminated with help of the identities: $\{\tau_z,\hat{g}^{R/A}_2\}\pm (\hat{g}^{R/A}_1)^2=0$ coming from the normalization condition $\check{g}^2=1$.) The function $\hat{f}_2$ vanishes at the boundaries with the leads and solves the differential equation:
\begin{eqnarray}
-2iD\nabla^2\hat{f}_2+2[J \sigma_z,\hat{f}_2]
=-iD\left(
\nabla\hat{g}^R_1\hat{g}^R_1 \nabla\hat{f}_0
\right.
\nonumber 
\\
\left.
+\nabla\hat{f}_0\hat{g}^A_1\nabla \hat{g}^A_1
+\nabla(\hat{g}^R_1\nabla\hat{f}_0\hat{g}^A_1)
\right).
\label{eq:f2}
\end{eqnarray}
Close to $T_c$, the Josephson relation remains sinusoidal, $I=I_c \sin\chi$, and the critical current $I_c$ can be obtained by evaluating Eq.~(\ref{eq:current2}). 

\begin{figure}
\includegraphics*[width=0.9 \columnwidth]{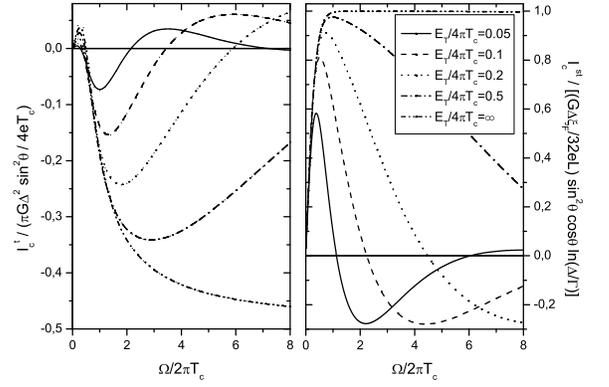}
\caption{Precession frequency dependence of long range (left) and ``anomalous'' (right) contributions to the critical current for ferromagnetic Josephson junctions with different lengths.}
\label{fig2}
\end{figure}

Before proceeding, let us make simplifying assumptions. In usual ferromagnets, $h_F$ would exceed $\Omega$ and $T_c$ by several orders of magnitude. Therefore, $\alpha\simeq \theta$, $J\simeq h_F$ and $p_-\simeq q$. We also assume that the junction is long ($L\gg\xi_F$). Thus, we discard the exponentially small terms contributing to $I_c$ and we find that it has two main terms: $I_c=I_c^t+I_c^{st}$. The first one comes from the long range triplet proximity effect (component $\bar{g}_{1x}$), only:
\begin{eqnarray}
I_c^t&=&-\frac{G\sin^2\theta}{2e}\int d\eps f_+\Im\left[(F_-^R)^2\frac{kL}{\sinh kL}\right]
\nonumber  \\
&=&
-\frac{G\pi T \Delta^2\sin^2\theta}{e}\Re\sum_{\omega>0}
\left(\frac{1}{\omega+i \Omega}-\frac{1}{\omega}\right)^2
\frac{\kappa_\omega L}{\sinh \kappa_\omega L},
\nonumber  \\
\label{eq:res1}
\end{eqnarray}
where $\omega=(2n+1)\pi T$ are Matsubara frequencies and $\kappa_\omega=\sqrt{2(\omega+i\Omega/2)/D}$. In particular, we get 
\be
I_c^t/I_{c0}\simeq
\left\{
\begin{array}{lcl}
\frac{1}{24}(\Omega/T_c)^2\sin^2\theta & \text{if} &  \Omega\ll T_c\ll E_T, \\
-\frac{1}{2}\sin^2\theta & \text{if} &   T_c\ll \Omega \ll E_T.
\end{array}
\right.
\ee
Here, $I_{c0}=\pi G \Delta^2/4eT_c$ is the critical current for a S/N/S junction with $T_c\ll E_T$ at temperatures close to $T_c$. We notice that $I_c^t$ vanishes at $\Omega=\pi T_c$. At larger frequency, it changes its sign and a $\pi$-coupling between the superconducting leads is realized. When $\Omega \sim E_T$, we also find an oscillatory behaviour of the frequency dependence of $I_c^t$ on the scale of $E_T$ (see fig.~\ref{fig2}) with an alternation of $0$- and $\pi$-couplings.

In contrast to ferromagnetic Josephson junctions with spatial variation of the magnetization and S/HM/S junctions, there arises a mixing of short range and long range proximity effects through the non-equilibrium spin-dependent terms of the distribution function matrix in the system studied here. Most of the terms in Eq.~(\ref{eq:current2}) reflecting such mixing are suppressed by the small factor $\xi_F/L$, when compared to $I_c^t$. However, the contribution 
\be
I_c^{st}=\frac{G\xi_F\sin^2\theta\cos\theta}{8eL}\int d\eps f_-|F^R_-|^2\Re\frac{kL}{\sinh kL}
\label{eq:res3}
\ee
coming from the ``anomalous term'' $\propto \hat{g}^R_1\nabla \hat{f}_0\hat{g}^A_1$ in Eqs.~(\ref{eq:current2})-(\ref{eq:f2}) needs more care. Indeed, for this term, a finite depairing parameter $\Gamma$ is necessary to regularize the otherwise diverging integral over the energies. Assuming $\Gamma\ll\Delta(T)$ and $\Omega\ll E_T$, we get in the logarithmic approximation:
\begin{equation}
\label{eq:res4}
I_c^{st}\simeq \frac{ G \Delta }{32 e} 
\left( \frac{\xi_F}{L} \right)
\sin^2\theta\cos\theta
\ln\frac{\Delta}{\Gamma}\tanh\frac{\Omega}{T_c}.
\end{equation}  
This term vanishes like $(T_c-T)^{1/2}\ln(T_c-T)$ close to $T_c$ . Such non analytic temperature dependence was already met in the field of transport in N/S junctions \cite{anomalous}. When $\Omega\sim E_T$, we also get from Eq.~(\ref{eq:res3}) an oscillatory frequency dependence of $I_c^{st}$ (see fig.~\ref{fig2}).

Let us now compare both main contributions (\ref{eq:res1}) and (\ref{eq:res3}) to $I_c$. They have quite different dependences on the device parameters. At low precession frequency compared to $T_c$, $I_c^{st}$ scales linearly with $\Omega$ and dominates over $I_c^{t}$ which scales quadratically. On the other hand, at large frequency, $I_c^{t}$ takes over if $\Delta/T_c \gtrsim (\xi_F/L)\ln(\Delta/\Gamma)$. At low temperatures, we speculate that the above expressions still hold qualitatively. As the ratio $\Delta/T_c$ becomes of the order of unity, the critical current at $L\gg\xi_F$ is dominated by the long range triplet component only. 

Simplifying assumptions have been made to get analytical expressions for the current. In principle, more realistic conditions could also be studied. The Usadel equation (\ref{eq:usadel2}) may be solved numerically to get the current at all temperatures. Spin-flip diffusion in the ferromagnet and tunnel barriers at the F/S interfaces could easily be incorporated in the theory. On the other hand, a finite spin polarization in the ferromagnet, as well as a mean free path comparable with the coherence length would require to go beyond the quasiclassical theory for diffusive metals. Nevertheless, we expect that Eqs.~(\ref{eq:res1})-(\ref{eq:res4}) hold qualitatively beyond their strict range of validity. Thus, our findings are relevant to the study of F/S junctions with conventional metals. The quasiclassical theory has proved its usefulness for the prediction and quantitative analysis of several experiments in these systems \cite{buzdin}.

The dynamically induced long range proximity effect studied in this work is expected to have special properties. In particular, it would be of interest to characterize the current-voltage characteristics in S/F and S/F/S junctions. We may expect an a.c. current response at frequencies mixing $\Omega$ and the Josephson frequency. We note this has been studied in the context of tunneling transport through a single magnetic impurity placed between two superconductors \cite{balatsky} as well as S/HM/S junctions with extended interfaces \cite{maekawa2}.

The long range Josephson current may be observed by performing FMR experiments in the microwave regime \cite{maekawa} in a planar S/F/S junction. This geometry was used in N/F/N junctions \cite{costache06,moriyama08} to detect electrically the spin pumping \cite{brataas} due to a precessing magnetization. A large resonance frequency, such that $\hbar\Omega_0\sim k_BT_c$ ($1\text{K}$ corresponds to $30\text{GHz}$), can be reached even by applying a moderate d.c. magnetic field $H_0$ along the plane of the ferromagnet \cite{kittel}. A small transverse r.f. field, $H_\text{rf}$, is used to induce the magnetization precession. The tilt angle $\theta$ strongly depends on the precession frequency. At resonance, $\theta$ can be estimated from the Landau-Lifshitz equations: $\theta\sim \mu H_{rf}/a \hbar \Omega_0$, where $\mu$ is the Bohr magneton and $a$ is the Gilbert damping parameter. In soft ferromagnets, a quite large $\theta\simeq 15^o$ could be obtained. Away from resonance, $\theta$ is much reduced. This would induce a strong frequency dependence, {\it via} angle $\theta$, in the expressions for the critical current derived above. We note that FMR was recently performed in an Nb/Permalloy bilayer \cite{aarts}. Below the superconducting critical temperature, the reduction of the resonance width was observed. It was attributed to an efficient proximity effect leading to the reduced efficiency of spin-flip processes below $T_c$ \cite{morten-belzig}. In particular, this shows that the proximity effect still persists at FMR. If the same conditions can be met in an S/F/S junction, we expect that the large critical current predicted in this work will be well measurable. 

In conclusion, we have proposed a model of ferromagnetic Josephson junction with precessing magnetization where a large current flows thanks to a dynamically induced long range proximity effect. The dependence of the current with the precession frequency shows an oscillatory behavior. We have discussed the conditions for this effect to be observed under the condition of ferromagnetic resonance in superconductor/ferromagnet devices with conventional metals.

The work presented above has benefited from discussions with S. Teber, R. M\'elin and D. Feinberg on a related problem, as well as with A. Buzdin, M. Aprili and T. Champel. Support from ANR-07-NANO011 ELEC-EPR is acknowledged.


\begin{thebibliography}{99}

\bibitem{buzdin}
	A. I. Buzdin, Rev. Mod. Phys. \textbf{77}, 935 (2005).

\bibitem{bergeret-rmp}
	A. Kadigrobov, R. I. Shekhter, and M. Jonson, Europhys. Lett. \textbf{54}, 394 (2001);
F. S. Bergeret, A. F. Volkov, and K. B. Efetov, Phys. Rev. Lett. \textbf{86}, 4096 (2001); Rev. Mod. Phys. \textbf{77}, 1321 (2005).
	
\bibitem{eschrig}
	M. Eschrig, J. Kopu, J. C. Cuevas, and G. Sch\"on, Phys. Rev. Lett. \textbf{90}, 137003 (2003);  Y. Asano, Y. Tanaka, A. A. Golubov, Phys. Rev. Lett. \textbf{98}, 107002 (2007); M. Eschrig and T. L\"ofwander, Nature Physics \textbf{4}, 138 (2008).
	
	
\bibitem{keizer}
	 R. S. Keizer, S. T. B. Goennenwein, T. M. Klapwijk, G. Miao,
G. Xiao, A. Gupta, Nature \textbf{439}, 825 (2006).
	
\bibitem{sosnin}
	I. Sosnin, H. Cho, V. T. Petrashov, and A. F. Volkov, Phys. Rev. Lett. \textbf{96}, 157002 (2006).
	
\bibitem{houzet-buzdin}
V. Braude and Yu. V. Nazarov, Phys. Rev. Lett. \textbf{98}, 077003 (2007)  ; 
M. Houzet and A. I. Buzdin, Phys. Rev. B \textbf{76}, 060504 (2007). 

\bibitem{brataas}
	Y. Tserkovnyak, A. Brataas, G. E. W. Bauer, and B. I. Halperin, Rev. Mod. Phys. \textbf{77}, 1375 (2005).

\bibitem{maekawa}
	S. Takahashi, S. Hikino, M. Mori, J. Martinek, and S. Maekawa, Phys. Rev. Lett. \textbf{99}, 057003 (2007).

\bibitem{usadel}
      K. D. Usadel, Phys. Rev. Lett. \textbf{25}, 507 (1970).

\bibitem{larkin}
	  A. I. Larkin and Yu. N. Ovchinnikov, in {\it Nonequilibrium superconductivity} (Eds. D. N. Langenberg and A.I. Larkin),   Elsevier Science Publishers (1986).

\bibitem{cuevas}
J. C. Cuevas, J. Hammer, J. Kopu, J. K. Viljas, and M. Eschrig, Phys. Rev. B \textbf{73}, 184505 (2006).



\bibitem{note}
	The retarded anomalous function reduces to $F^R(\eps)=i\Delta/(\eps+i0^+)$ when $\Delta,\Gamma \ll |\eps|$. Below, we will see a circumstance where the full denominator should be kept, still assuming $\Delta,\Gamma \ll T_c$.

\bibitem{anomalous}
    R. Seviour, C. J. Lambert, and A. F. Volkov, Phys. Rev. B \textbf{59}, 6031 (1999).

	
\bibitem{balatsky}
Jian-Xin Zhu and A. V. Balatsky, 
Phys. Rev. B \textbf{67}, 174505 (2003).

	
\bibitem{maekawa2}
 S. Hikino, M. Mori, S. Takahashi, S. Maekawa,  arXiv:0802.1755.
	
\bibitem{kittel}
C. Kittel, {\it Introduction to solid state physics} (John Wiley, 2005).
	
\bibitem{costache06}
M. V. Costache, M. Sladkov, S. M. Watts, C. H. van der Wal, and B. J. van Wees,
Phys. Rev. Lett.  \textbf{97}, 216603 (2006).	


\bibitem{moriyama08}
T. Moriyama, R. Cao, X. Fan, G. Xuan, B. K. Nikolic, Y. Tserkovnyak, J. Kolodzey, and John Q. Xiao, Phys. Rev. Lett. 100, 067602 (2008). 
	
	
\bibitem{aarts}
C. Bell, S. Milikisyants, M. Huber, and J. Aarts, 
Phys. Rev. Lett. \textbf{100}, 047002 (2008).

\bibitem{morten-belzig}
 J. P. Morten, A. Brataas, G. E. W. Bauer, W. Belzig, Y. Tserkovnyak, arXiv:0712.2814.
 
\end{thebibliography}
\end{document}